\documentclass[aps,prb,twocolumn,groupedaddress]{revtex4}
\usepackage{graphicx}
\begin{document}
%\preprint{}


\title{The Mott State and Superconductivity in Face-Centred Cubic Structured Cs$_3$C$_{60}$: A $^{133}$Cs-Nuclear Magnetic Resonance Study under Pressure}

\author{Shinji Kawasaki$^1$}
\author{Junji Fukui$^1$}
\author{Takeshi Motoyama$^1$}
\author{Yuta Suzuki$^1$}
\author{Seiji Shibasaki$^1$}
\author{Guo-qing Zheng$^{1,2}$}

\affiliation{$^1$Department of Physics, Faculty of Science, Okayama University, Okayama 700-8530, Japan} 
\affiliation{$^3$Institute of Physics and Beijing National Laboratory for Condensed Matter Physics, Chinese Academy of Sciences, Beijing 100190, China}


%\author{}
%\email[zheng@psun.phys.okayama-u.ac.jp]{Your e-mail address}
%\thanks{}
%\affiliation{}


%\date{\today}

\begin{abstract}
Over the past 20 years, fullerides have been studied as the source of high-transition-temperature ($T_c$) superconductivity except for copper oxides. The recent finding of the Mott insulating state right beside superconductivity in Cs$_3$C$_{60}$ has suggested that magnetism helps raise $T_c$ even in fullerides as in heavy-fermion compounds, high-$T_c$ copper oxides, two-dimensional organic conductors, and iron pnictides. Namely, one tends to think that the link between Mott insulator and superconductivity takes place in fullerides, which can give rise to the mechanism beyond the Bardeen-Cooper-Schrieffer framework. However, the relationship between the Mott state and the superconductivity in Cs$_3$C$_{60}$ is still under debate. By nuclear magnetic resonance measurements under pressure, we find that the magnetism and superconductivity in Cs$_3$C$_{60}$ are competing orders.  Different from previous reports, the phase separation of Cs$_3$C$_{60}$ crystals into the Mott and metallic states allows us to systematically study the evolution of the ground state under pressure. Our careful experiments have found that the prevention of a magnetic order is rather essential for the superconductivity in face-centred cubic Cs$_3$C$_{60}$, which presents a basic strategy for finding still higher $T_c$ in this system.
\end{abstract}

\maketitle

\section{Introduction}

The alkali-doped fullerides $A_3$C$_{60}$ ($A$: alkali ion) have been studied as the source of high-transition-temperature ($T_c$) superconductivity\cite{Capone}. The highest $T_c$ of 38 K has been reported in the A15 phase of Cs$_3$C$_{60}$ under pressure\cite{Ganin,Takabayashi,Ganin2}. The reported superconducting dome beside a Mott insulating phase\cite{Ganin,Takabayashi,Ganin2} is similar to that in heavy-fermion compounds\cite{Mathur}, high $T_c$ copper oxides\cite{PALee}, two-dimensional organic conductors\cite{Kanoda1,Kanoda2}, and iron pnictides\cite{Kamihara,Oka}. Furthermore, the coexistence of antiferromagnetism and superconductivity has been suggested\cite{Takabayashi}. In order for superconductivity to survive and even to be enhanced in the vicinity of a Mott state, new mechanisms beyond the Bardeen-Cooper-Schrieffer (BCS) framework have been expected\cite{Capone,Ganin,Takabayashi,Ganin2,Iwasa,IharaPRL,IharaEPL}.

Among its polymorphism, the face-centred-cubic (fcc) $A_3$C$_{60}$ also shows high-$T_c$ superconductivity. Its $T_c$ increases with increasing lattice constant by replacing the $A$ with a larger alkali ion\cite{Hebard,Fleming,Tanigaki}. Replacing the alkali ion K with Rb significantly increases $T_c$ from 20 to 30 K\cite{Fleming}.  Furthermore, replacing Rb with Cs resulted in $T_c$ = 31.3 K for Rb$_2$CsC$_{60}$ \cite{Fleming} and in $T_c$ = 33 K for RbCs$_2$C$_{60}$ \cite{Tanigaki}. This relationship between $T_c$ and the lattice parameter suggests that the superconductivity is due to a conventional electron-phonon interaction because the decrease in the extent of overlapping of the molecular wave function drives the density of states, $N (E_F)$, up, and thus $T_c$ increases\cite{Stenger,Pennington}. However, since the large C$_{60}$-C$_{60}$ distance causes the localization of electrons on C$_{60}$ balls, the end member of fcc Cs$_3$C$_{60}$ becomes a Mott insulator with $T_N$ $\sim$ 2.5 K\cite{Ganin2}. $T_N$ is one order of magnitude smaller and the observed internal magnetic field is much smaller than those of A15 Cs$_3$C$_{60}$ owing to geometrical spin frustration\cite{Ganin,Takabayashi,Ganin2}.  Notably, the highest $T_c$ of 35 K among fcc $A_3$C$_{60}$'s is found after suppressing this Mott state under pressure\cite{Ganin2}. Thus, whether the metal-insulator transition plays any role in inducing high-$T_c$ superconductivity has become an issue in fullerides \cite{Capone,Ganin,Takabayashi,Ganin2,Iwasa,IharaPRL,IharaEPL}. The previous nuclear magnetic resonance (NMR) experiment suggested that the end point of metal-insulator transition exists above the superconducting dome, indicating that the electronic correlations leading to the Mott state also play a role in having a maximal $T_c$ adjacent to the Mott phase in Cs$_3$C$_{60}$ under pressure\cite{IharaEPL}.
 
In this paper, we report that the Mott state and superconductivity of fcc Cs$_3$C$_{60}$ are competing orders. Our careful $^{133}$Cs-NMR spectrum and nuclear spin-lattice-relaxation time ($T_1$) measurements revealed that there are two different electronic phases for fcc Cs$_3$C$_{60}$. We find that the Mott state is robust against pressure and that the pressure-temperature phase diagram reported thus far\cite{Ganin,Ganin2,IharaEPL} is difficult to understand only by the pressure effect on the Mott state. The other seed for superconductivity, namely, paramagnetic metal, is hidden in the background of the Mott state, which realizes the phase separation of the two orders of antiferromagnetism and superconductivity in the low-pressure region. For comparison, with a previous structural study under pressure\cite{Ganin2}, we present a new phase diagram of temperature versus unit-cell volume around the Mott insulator-to-metal transition, which unambiguously indicates that the Mott state is unfavourable for forming Cooper pairs in Cs$_3$C$_{60}$. 

\section{Experimental Procedure}
fcc Cs$_3$C$_{60}$ crystals were synthesized by a solution process with CH$_3$NH$_2$ at low temperatures\cite{Ganin}. Nominal amounts of Cs and C$_{60}$ were introduced into glass tubes in an Ar glove box (O$_2$ $<$ 0.1 ppm and H$_2$O $<$ 0.1 ppm), and mixed in CH$_3$NH$_2$ solution. Powder X-ray diffraction patterns were measured with RIGAKU TTR-III. The lattice parameter of 14.752 $\AA$ at room temperature, which is consistent with a previous report of 14.761 $\AA$\cite{Ganin}, was determined by Rietveld refinement with the GSAS program\cite{Larson}. The fraction of the fcc polymorphic phase used in this experiment is 70 \% (fcc : A15 : bco = 0.70 : 0.23 : 0.07).  Here, bco denotes the body-centred-orthorhombic structure\cite{Ganin,IharaPRL,Jeglic}. This fraction of the fcc phase is comparable to previous reports\cite{Ganin,IharaPRL}. The magnetization $M$ was measured using a SQUID magnetometer (Quantum Design MPMS2) in the temperature range of $T$ $>$ 2 K.

BeCu and NiCrAl/BeCu piston-cylinder-type pressure cells were used to obtain the temperature dependence of magnetization and to perform NMR measurement under pressure, respectively. Daphne oil 7373 is used as a pressure-transmitting medium\cite{Murata}. The samples are paraffin-embedded to avoid possible degradation by Daphne oil under pressure.  The $T_c$ of Sn was monitored to determine pressure at low temperatures.

NMR measurement was carried out using a phase-coherent spectrometer. The usual Hahn-echo method ($\pi/2$$-$$\tau$$-$$\pi$), which reduces the A15 contribution to the signal by about a factor of 3, is successfully used to obtain only the fcc phase among its polymorphism\cite{Ganin,IharaPRL} [see Fig. 1(a)]. The NMR spectrum is obtained by the Fourier transform of spin echo with a typical pulse sequence of 5$\mu$s$-$ $\tau$ $-$10$\mu$s and by plotting the spin echo intensity while sweeping the external magnetic field at a fixed frequency with typical pulse sequence of 60$\mu$s $-$ $\tau$ $-$ 120$\mu$s. The nuclear recovery curves for obtaining $T_1$ were measured using the first pulse length of about 10 $\mu$s.

\section{Experimental Results}
\subsection{$^{133}$Cs-NMR spectrum and $T_1$}
Figure 1(a) shows the $^{133}$Cs-NMR spectrum measured at $T$ = 200 K and $P$ = 0. Crystallographically, fcc Cs$_3$C$_{60}$ has two $^{133}$Cs sites. One is surrounded by octahedral C$_{60}$ balls [Cs$_3$C$_{60}$ (O)], the other is surrounded by tetrahedral C$_{60}$ balls [Cs$_3$C$_{60}$ (T)]\cite{Ganin2}. Fundamentally, both sites are in a single electronic state, i.e., the temperature dependences of $T_1$ for both sites are the same\cite{Stenger}. On the other hand, it has been known that the other site (T$^\prime$) in fcc $A_3$C$_{60}$ appears below room temperature\cite{Ganin2,IharaPRL,Pennington,Walstedt,Gorny,Maniwa,Alloul,Zimmer,Skadchenko}. It has been suggested that it comes from some crystal distortions such as displacements or vacancies of $A$ ions, a Jahn-Tellar distortion of the C$_{60}$ ion,  and a deviation from the expected C$_{60}$ orientation in fcc $A_3$C$_{60}$ crystals\cite{Walstedt,Gorny,Apostol}. Furthermore, in the Na$_2A$C$_{60}$ compound, which has a simple cubic structure, the merohedral disorder is proposed as the origin of the T$^\prime$ site\cite{Matus}.  As shown in Fig. 1(a), the sharp peak next to the peak T arises from this T$^\prime$ site [Cs$_3$C$_{60}$ (T$^\prime$)].  For our Cs$_3$C$_{60}$ crystal, it is estimated from Fig. 1(a) that the volume fraction ratio of the T site to the T$^\prime$ site is 5.8 : 1 and the volume fraction of the T$^\prime$ site in the entire spectrum is estimated to be 13\%, which is consistent with the results of previous NMR studies of fcc $A_3$C$_{60}$ \cite{Ganin2,IharaPRL,Pennington,Walstedt,Gorny,Apostol,Maniwa,Alloul,Zimmer,Skadchenko}.  Note that the origin of the T$^\prime$ site for fcc $A_3$C$_{60}$ remains unresolved\cite{Ganin2,IharaPRL,Pennington,Walstedt,Gorny,Apostol,Maniwa,Alloul,Zimmer,Skadchenko}. However, it has been suggested by double resonance measurement and widely accepted that the T$^\prime$ site forms clusters, which exist randomly in $A_3$C$_{60}$ crystals without site-to-site correlation.\cite{Walstedt,PenningtonSEDOR} 

\begin{figure}
\begin{center}
\includegraphics[width=8.5cm]{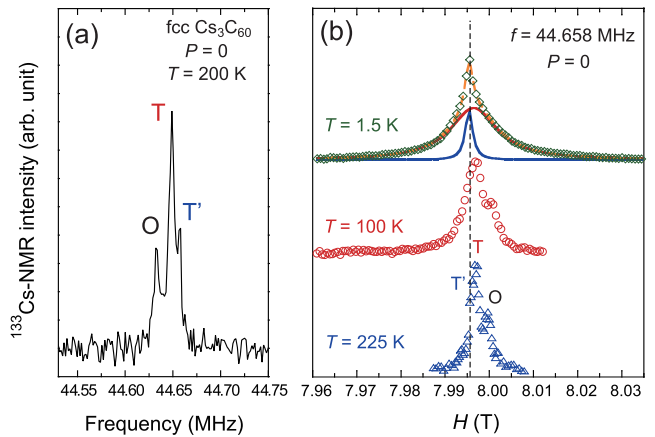}
\end{center}
\caption{(a) $^{133}$Cs-NMR ($^{133}\gamma$ = 5.5844 MHz/T) spectrum of fcc Cs$_3$C$_{60}$ measured at $H$ = 8.0 T. 
(b) Temperature dependence of Cs-NMR spectrum obtained by sweeping the external magnetic field at the fixed frequency of 44.658 MHz. At $T$ = 1.5 K, the fitting curves are shown together. The dotted curve is the result of Lorentzian fitting, which consists of broad and sharp peaks. This unambiguously indicates that there exist two fcc Cs$_3$C$_{60}$ phases with and without internal magnetic field (Mott state and paramagnetic metal). The dotted line indicates the position of the T$^\prime$ peak.}
\label{f1}
\end{figure}

Figure 1(b) shows the temperature dependence of the $^{133}$Cs-NMR spectrum obtained by sweeping the external magnetic field at a fixed frequency of $f$ = 44.658 MHz.  As seen in the figure, the spectrum obtained at $T$ = 1.5 K consists of two peaks, a sharp and a broad peak, indicating that there are two electronically different phases with and without an internal magnetic field (Mott and paramagnetic states) as the ground state, as will be discussed later. Notably, as shown by the dotted straight line, the peak position of the sharp peak is consistent with that for the T$^{\prime}$ site at high temperatures. In addition, the volume fraction ($\sim$15 \%) of this sharp peak in the entire spectrum at $T$ = 1.5 K is also consistent with that for the T$^\prime$ peak at high temperatures. Thus, it is indicated that these broad and sharp peaks originate from the T and T$^\prime$ sites, respectively.

\begin{figure}
\begin{center}
\includegraphics[width=8cm]{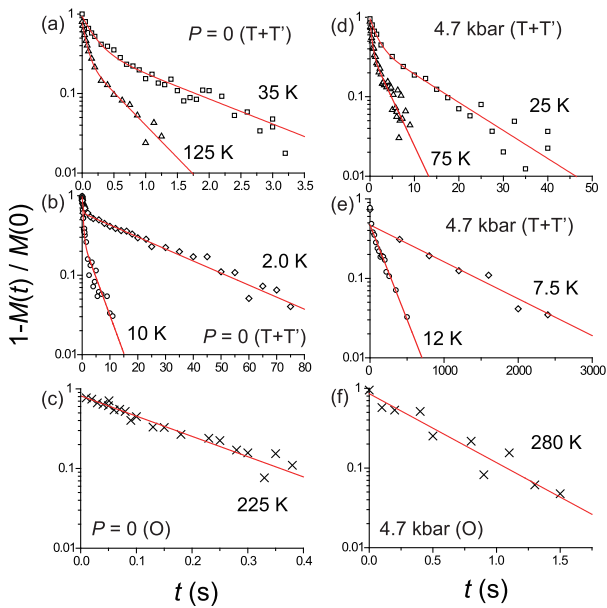}
\end{center}
\caption{Typical datasets of nuclear magnetization decay curves for obtaining $T_1$ for T + T$^\prime$ sites (a), (b) [(d),(e)] and O site (c) [(f)] at $P$ = 0 kbar ($P$ = 4.7 kbar), respectively. Solid curves are theoretical fitting curves assuming two different $T_1$s as parameters as 1$-M(t)$/$M_0$ = $\alpha$ exp$(-t/T_1^T$)+(1$-\alpha$)exp$(-t/T_1^{T'}$).  The contribution of $T_1$ to the O site to $T_1$ for T + T$^{\prime}$ sites is negligible.  The $T_1$ for the O site can be fitted by a single component as 1$-M(t)/M_0$ $=$ $\alpha$ exp$(-t/T_1^O)$ (c), (f). Well below $T_c$ ($T$ = 12 K and 7.5 K), only the long component $T_1$ is shown (e).
}
\label{f1}
\end{figure}
 
To elucidate the electronic property for fcc Cs$_3$C$_{60}$, we measured $^{133}$Cs $T_1$ systematically. For fcc Cs$_3$C$_{60}$, since there is no quadrupole interaction at the Cs site owing to highly symmetrical C$_{60}$ balls surrounding Cs ions\cite{Ganin2,IharaPRL,Pennington}, the nuclear magnetization decay curve for obtaining $T_1$ for the $^{133}$Cs nucleus is a single-exponential function\cite{Ganin2,IharaPRL,Pennington}, 1$-M(t)/M_0$ = exp$(-t/T_1)$, where $M_0$ and $M(t)$ are the nuclear magnetizations in the thermal equilibrium and at a time $t$ after a saturation pulse. Figures 2(a), 2(b), 2(d), and 2(e) show typical datasets of the nuclear magnetization decay curves obtained through the integration of both the T and T$^\prime$ peaks.  
Figures 2(c) and 2(f) show typical datasets of the nuclear magnetization decay curves obtained through the integration of the O peak. At high temperatures, since the O peak in the spectrum is well separated from the T and T$^{\prime}$ peaks, we can obtain $T_1$ for Cs$_3$C$_{60}$ (O) from a single-component decay curve as 1$-M(t)/M_0$ = $\alpha$ exp$(-t/T_1^O)$.  The absolute value of $T_1^O$ is consistent with previous reports\cite{Ganin,IharaPRL}.

On the other hand, for the T and T$^\prime$ peaks, we observed two $T_1$ components, i.e., a short one and a long one.  The solid curves are fittings consisting of a short $T_1$ ($T_1^S$) and a long $T_1$ ($T_1^L$) as $1-M(t)/M_0$ = $\alpha$ exp$(-t/T_1^S)$ + (1$-\alpha$) exp$(-t/T_1^L)$. Here, the absolute value of $T_1^S$ is consistent with previously reported $T_1$ for the T site\cite{Ganin,IharaPRL}.  However, $T_1^L$ has not been reported for Cs$_3$C$_{60}$, although, most probably, it originates from the T$^{\prime}$ site.  As will be discussed later, these two $T_1$s show completely different temperature dependences, indicating that the phase separation into two electronic states occurs in our fcc Cs$_3$C$_{60}$ crystal, which is consistent with the NMR spectrum [see Fig. 1(b)].   
 
The present results are different from the results of previous NMR studies on Cs$_3$C$_{60}$\cite{Ganin,IharaPRL}. One reports a shorter $T_1$ for the T$^\prime$ site than for the T site, and the temperature dependences of $T_1$ for both sites are similar\cite{Ganin}. The other report suggested that $T_1$'s for the T and T$^\prime$ sites are the same\cite{IharaPrivate}. Although the origin of these differences is still unknown, from the temperature dependences of the Cs-NMR spectrum and $T_1$,  we conclude that our Cs$_3$C$_{60}$ crystal has different electronic states in the T and T$^\prime$ sites as the ground state.  We just speculate that the T$^{\prime}$ site in our Cs$_3$C$_{60}$ crystal does not exist as clusters but as domains to form its own phase. A phase separation is often seen in strongly correlated electron systems such as manganese oxides\cite{Dagotto}, and its mechanism might share some similarities across different classes of these materials.  
 Here, we denote these two different phases for the T and T$^\prime$ sites as the Cs$_3$C$_{60}$ (T) and Cs$_3$C$_{60}$ (T$^\prime$) phases, respectively. As shown in the recovery curves, this two phases feature does not change under pressure.

\begin{figure}
\begin{center}
\includegraphics[width=7.5cm]{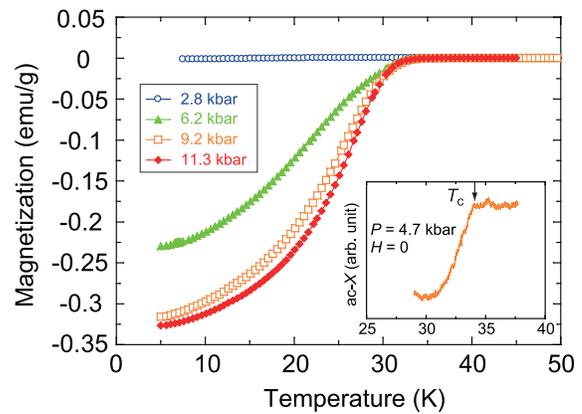}
\end{center}
\caption{Temperature and pressure dependences of magnetization for fcc Cs$_3$C$_{60}$. The $T_c$'s (superconducting shielding fractions) of 33.8 K (33.1\%), 35.6 K (45.5\%), and 34.5 K (47.1\%) are obtained at $P$ = 6.2, 9.2, and 11.3 kbar, respectively. The inset shows the temperature dependence of the ac susceptibility deduced from the detuning of the NMR tank circuit using an $in$ $situ$ NMR coil at $P$ = 4.7 kbar under a zero magnetic field. ($T_c$ = 34.0 K).  }
\label{f1}
\end{figure}

\subsection{$P$-induced superconductivity in fcc Cs$_3$C$_{60}$}
In this section, we present the results of the pressure-induced superconductivity of the fcc Cs$_3$C$_{60}$ crystal. Figure 3 shows the temperature dependence of magnetization. As seen in Fig. 3, the superconducting shielding fraction depends on pressure. This is consistent with previous results\cite{Ganin2,IharaPRL}. From Fig. 3, it is suggested that the superconductivity in full volume is realized only above $P$ $\sim$ 9 kbar. This is in contrast with that observed in heavy-fermion compounds \cite{SK} or iron-pnictides \cite{ZLi} where  bulk superconductivity has been observed even in the coexisting state with antiferromagnetism.

\subsection{$T$ and $P$ dependences of 1/$T_1T$}
Next, we show the evolution of the electronic properties of the Cs$_3$C$_{60}$ (T) and Cs$_3$C$_{60}$ (T$^\prime$) phases and its relation to the superconductivity under pressure. Figure 4 shows the temperature dependences of nuclear spin-lattice relaxation rate (1/$T_1$) divided by temperature (1/$T_1T$) for the Cs$_3$C$_{60}$ (T\&O) and Cs$_3$C$_{60}$ (T$^\prime$) phases measured at $P$ = 0 and 4.7 kbar, respectively. In general, 1/$T_1T$ is proportional to  electron spin susceptibility through the hyperfine coupling constant $A_{\vec{q}}$ as
\begin{equation}
\frac{1}{T_1T} \propto \sum_{\vec{q}} \left | A_{\vec{q}} \right |^2 \frac{\chi ^{\prime\prime}(\vec{q}, \omega_0)}{\omega_0},
\end{equation}
where $\omega_0$  is the NMR frequency. In fullerides, the dynamic spin susceptibility $\chi^{\prime\prime}(\vec{q}, \omega)$  is caused by the antiferromagnetic fluctuation with a finite $\vec{q}$  due to the localized electron spins on C$_{60}$ molecules\cite{Pennington}. Furthermore, if the lattice parameter decreases, then the overlapping of the C$_{60}$ molecular wave function increases and $A_3$C$_{60}$ becomes a paramagnetic metal\cite{Pennington}. In such a case, since $\chi ^{\prime\prime}(\vec{q}, \omega)$  is proportional to the square of the density of states, $N(E_F)$, at the Fermi level and $A_{\vec{q}}$ is $\vec{q}$-independent, eq. (1) yields the relation $\frac{1}{T_1T}$ $\propto$ $\left | A_{\vec{q}} \right |^2N(E_F)^2$  (Korringa-law). In particular, in fcc $A_3$C$_{60}$ crystals, 1/$T_1T$'s for the T and O sites are connected through each hyperfine coupling constant\cite{Stenger} as

\begin{equation}
\left( \frac{1}{T_1T}\right)^{O} \propto \left | \frac{A^O_{\vec{q}}}{A^T_{\vec{q}}} \right |^2 \left( \frac{1}{T_1T}\right)^{T}.
\end{equation}

As seen in Fig. 4, 1/$T_1T$ for the Cs$_3$C$_{60}$ (T) phase at $P$ = 0 and 4.7 kbar increases with decreasing temperature. This indicates that the antiferromagnetic fluctuation develops as a result of the localization of electron spins on C$_{60}$ balls. At $P$ = 0, 1/$T_1T$ shows a peak at $T_N$ $\sim$ 2.5 K due to antiferromagnetic order. These results at $P$ = 0 are consistent with previous $\mu$SR measurements\cite{Ganin}. At $P$ = 4.7 kbar, we also observed an increase in 1/$T_1T$ down to $T$ = 1.5 K, suggesting that electron spins are still localized to lead to the Mott state. Notably, no superconducting transition is observed in the Cs$_3$C$_{60}$ (T) phase at $P$ = 4.7 kbar although the onset of pressure-induced superconductivity at $T_c$($P$) = 34.0 K is confirmed by ac susceptibility measurement using an $in$ $situ$ NMR coil (see Fig. 3 inset). This is one of the most important results of this study, indicating that $P_c$ for the Mott state is above $P >$ 4.7 kbar. 

On the other hand, we found a very different temperature dependence of 1/$T_1T$ for the Cs$_3$C$_{60}$ (T$^\prime$) phase. At $P$ = 0, 1/$T_1T$ satisfies the Korringa law at low temperature, indicating that the Cs$_3$C$_{60}$ (T$^\prime$) phase is a paramagnetic metal. These results are consistent with the Cs-NMR spectrum shown in Fig. 1(b).  At $T$ = 1.5 K, the Cs-NMR spectrum consists of a broad peak for Cs$_3$C$_{60}$ (T), which originates from the hyperfine field induced by the localized electron spins in the Mott state, and a sharp peak for Cs$_3$C$_{60}$ (T$^\prime$), where such a field is absent in the metallic state. Thus, it is evidenced that the electronic state for the Cs$_3$C$_{60}$ (T$^\prime$) phase is completely different from that for the Cs$_3$C$_{60}$ (T) phase, and that both phases are spatially separated. Since this separation between the Cs$_3$C$_{60}$ (T$^\prime$) and Cs$_3$C$_{60}$ (T) phases is realized within the same fcc Cs$_3$C$_{60}$ crystal, it is suggested that the origin of the T$^{\prime}$ site for our fcc Cs$_3$C$_{60}$ crystal is a positive chemical pressure (internal pressure, $P_{int}$) to delocalize electron spins on C$_{60}$ balls, which causes the metallic state for the Cs$_3$C$_{60}$ (T$^\prime$) phase.

At $P$ = 4.7 kbar, the superconducting transition is observed only in the Cs$_3$C$_{60}$ (T$^\prime$) phase by a substantial decrease in 1/$T_1T$ below $T$ $\sim$ 30 K, contrary to the fact that no signature of superconductivity is seen in 1/$T_1T$ for the Mott-insulating Cs$_3$C$_{60}$ (T) phase. This is the source of the pressure-induced superconductivity of Cs$_3$C$_{60}$ in the low-pressure region. This is consistent with the observation of the partial superconducting shielding fraction at $P$ $<$ 9 kbar. From these results, it is clear that the superconductivity for fcc Cs$_3$C$_{60}$ is induced only for the paramagnetic metal.  Notably, the temperature dependence of 1/$T_1T$ for the Cs$_3$C$_{60}$ (T$^\prime$) phase at $P$ = 4.7 kbar traces the 1/$T_1T$ of $^{133}$Cs-NMR for Rb$_2$CsC$_{60}$ (O)\cite{Stenger}. Namely, 1/$T_1$ decreases exponentially below $T_c$ and the Korringa law is established above $T_c$.  Since the lattice parameter of fcc Rb$_2$CsC$_{60}$ is smaller than that of fcc Cs$_3$C$_{60}$, it is confirmed, from the electronic point of view, that the pressure effect on fcc $A_3$C$_{60}$ and the variation in its lattice parameter are equivalent.
 
It has been suggested that the mechanism of superconductivity in Rb$_2$CsC$_{60}$ is a conventional BCS one\cite{Stenger}. The present result indicates that the mechanism of Cooper pair formation for fcc Cs$_3$C$_{60}$ under pressure is also in the framework of the BCS one. In such a case, localized electron spins on C$_{60}$ balls in the Mott state break Cooper pairs\cite{Abrikosov}. This also supports the scenario that the origin of the "coexistence" of a Mott insulator and superconductivity in fcc Cs$_3$C$_{60}$ is the phase separation of the two orders in real space. 

\begin{figure}
\begin{center}
\includegraphics[width=7cm]{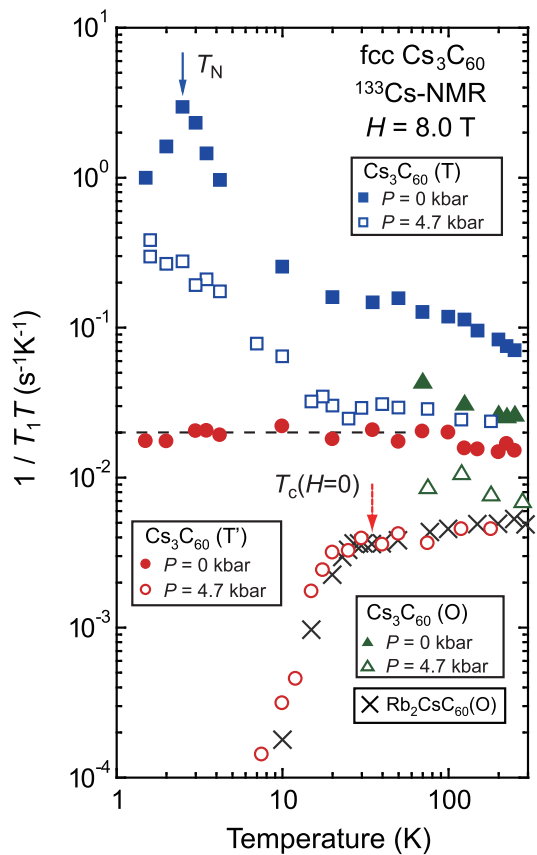}
\end{center}
\caption{Temperature and pressure dependences of $^{133}$(1/$T_1T$) for fcc Cs$_3$C$_{60}$ plotted on logarithmic scale. $T_1$'s for both Cs$_3$C$_{60}$ (T) and Cs$_3$C$_{60}$ (T$^\prime$) phases were measured in the range of 1.4 K $<$ $T$ $<$ 300 K. The $T_1$ for Cs$_3$C$_{60}$ (O) is  obtained above $T$ $>$ 70 K.  Solid (open) squares, triangles, and circles indicate 1/$T_1T$'s for Cs$_3$C$_{60}$ (T), Cs$_3$C$_{60}$ (O), and Cs$_3$C$_{60}$ (T$^\prime$) at $P$ = 0 ($P$ = 4.7 kbar), respectively. The solid and dotted arrows indicate $T_N$ and $T_c$ ($H$ = 0), respectively. $T_c$ ($H$ = 0) = 34.0 K is determined by ac-susceptibility measurement using an $in$ $situ$ NMR coil. The $^{133}$(1/$T_1T$) for Rb$_2$CsC$_{60}$ (O) (solid cross) is cited from the literature \cite{Stenger}. The dotted line is an eye guide indicating the Korringa law (1/$T_1T$ = constant.) at low temperature. }
\label{f1}
\end{figure}

\section{Discussion}
To explain the present results systematically, we suggest that the phase separation originates from the formation of domains of the Cs$_3$C$_{60}$ (T$^\prime$) phase due to local $P_{int}$ in fcc Cs$_3$C$_{60}$ crystals.  At $P$ = 0, paramagnetic domains of the Cs$_3$C$_{60}$ (T$^\prime$) phase are formed within the Mott-insulating Cs$_3$C$_{60}$ (T) phase. At $P$ = 4.7 kbar, as shown by the temperature dependence of 1/$T_1T$, only domains of the Cs$_3$C$_{60}$ (T$^\prime$) phase become superconducting and a partial shielding fraction is observed. Note that the superconducting coherence length for fcc Rb$_3$C$_{60}$ ($T_c$ = 31 K) is $\sim$19 $\AA$\cite{Bunter}. From this, it can be speculated that each domain will consist of more than 2 unit cells. Above $P >$ 4.7 kbar, an insulator-to-metal transition occurs in the Cs$_3$C$_{60}$ (T) phase. Following that, superconducting domain size naturally increases through the proximity effect at the boundary of superconducting domains and the metallic region and thus, the shielding fraction increases. Above $P >$ 9 kbar, pressure-induced superconductivity also occurs in the Cs$_3$C$_{60}$ (T) phase and the bulk superconductivity of the fcc Cs$_3$C$_{60}$ crystal sets in.  This scenario can explain the pressure dependence of the superconducting shielding fraction qualitatively and is in good agreement with NMR results as well.    
  
\begin{figure}
\begin{center}
\includegraphics[width=8cm]{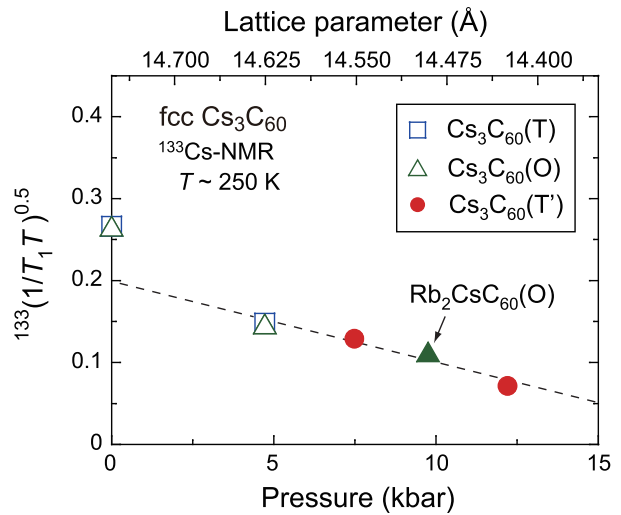}
\end{center}
\caption{Open squares indicate the pressure dependence of $^{133}$(1/$T_1T$)$^{0.5}$ for Cs$_3$C$_{60}$ (T). Open (solid) triangles indicate $^{133}$(1/$T_1T$)$^{0.5}$ for Cs$_3$C$_{60}$(O) [Rb$_2$CsC$_{60}$(O)]. 1/$T_1T$'s for O sites are normalized to that for the T site using the relation $\left(\sqrt{\frac{1}{T_1T}} \right)^O$ $\propto$  $\frac{A^O_{\vec{q}}}{A^T_{\vec{q}}}$ $\left(\sqrt{\frac{1}{T_1T}} \right)^T$ = $\frac{1}{1.63}$ $\left(\sqrt{\frac{1}{T_1T}} \right)^T$ . 1/$T_1T$ for Rb$_2$CsC$_{60}$ (O) is cited from the literature\cite{Stenger}. The normalized $^{133}$(1/$T_1T$)$^{0.5}$ for Rb$_2$CsC$_{60}$(O) corresponds to that for Cs$_3$C$_{60}$ (T) at $P$ = 9.7 kbar. Solid circles indicate $^{133}$(1/$T_1T$)$^{0.5}$ for Cs$_3$C$_{60}$ (T$^\prime$) corrected using the internal pressure of $P_{int}$ = 7.5 kbar. The lattice parameters at room temperature are estimated from the compressibility $\kappa$ = 0.0053/kbar for fcc Cs$_3$C$_{60}$ (ref. 4). The lattice parameter of 14.752 $\AA$ is used as the value at $P$ = 0 for Cs$_3$C$_{60}$ (T). The dotted line is an eye guide.  }
\label{f1}
\end{figure}

Finally, on the basis of the present results and considering the spatial phase separation due to $P_{int}$ in the fcc Cs$_3$C$_{60}$ crystal, we present a new phase diagram of fcc Cs$_3$C$_{60}$ using the unit cell volume ($\AA^3$) at room temperature. In Fig. 5, we plot the pressure dependences of $^{133}$(1/$T_1T$)$^{0.5}$ [$\propto$ $N(E_F)$  for paramagnetic metal] for the T, O, and T$^\prime$ sites of fcc Cs$_3$C$_{60}$ and for the O site of Rb$_2$CsC$_{60}$ (ref. 17). Here, the previous structural study has revealed that the unit cell volume for fcc Cs$_3$C$_{60}$ decreases linearly with the compressibility $\kappa$ = 0.0053/kbar at $P <$ 10 kbar\cite{Ganin}. From this relation, compared with the lattice parameter of 14.752 $\AA$ for our fcc Cs$_3$C$_{60}$ crystal at $P$ = 0, that the lattice parameter of 14.493 $\AA$ for Rb$_2$CsC$_{60}$ (ref. 15) corresponds to the lattice parameter for our fcc Cs$_3$C$_{60}$ at $P$ = 9.7 kbar. Meanwhile, using eq. (2), we obtained an exact match in the pressure dependence of $^{133}$(1/$T_1T$)$^{0.5}$ for the T and O sites. Notably, offsetting the Cs$_3$C$_{60}$ (T$^\prime$) data by $P_{int}$ = 7.5 kbar will make all data points fall on a straight line above $P >$ 5 kbar. This evidences that the Mott-insulating Cs$_3$C$_{60}$ (T) phase becomes metallic above $P >$ 5kbar, where $\left( \frac{1}{T_1T} \right)^{0.5}$ $\propto$ $N(E_F)$ $\propto$ $P$ holds. The deviation from a linear relation at $P$ = 0 is due to the development of antiferromagnetic fluctuations that lead to the Mott state. From Fig. 5 and using $\kappa$ = 0.0053/kbar, we estimate the lattice parameters of the Cs$_3$C$_{60}$ (T) and Cs$_3$C$_{60}$ (T$^\prime$) phases under pressure. The lattice parameter of the Cs$_3$C$_{60}$ (T) phase at $P$ = 4.7 kbar is 14.628 $\AA$. Since the Cs$_3$C$_{60}$ (T$^\prime$) phase at $P$ = 0 corresponds to the Cs$_3$C$_{60}$ (T) phase at $P$ = 7.5 kbar, the lattice parameter of the Cs$_3$C$_{60}$ (T$^\prime$) phase at $P$ = 0 ($P$ = 4.7 kbar) is estimated to be 14.554 $\AA$ (14.426 $\AA$).

\begin{figure}
\begin{center}
\includegraphics[width=8.5cm]{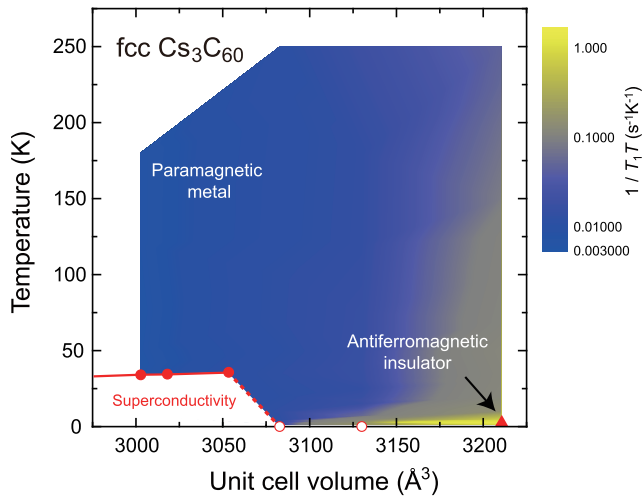}
\end{center}
\caption{Unit cell volume vs temperature phase diagram for fcc Cs$_3$C$_{60}$ plotted using the temperature dependence of 1/$T_1T$. Unit cell volume is calculated using the lattice parameter at room temperature. The color coding (bar on right) refers to 1/$T_1T$ on the logarithmic scale. The solid circles and triangle indicate $T_c$ and $T_N$, respectively. The open circles plotted at $T$ = 1.4 K indicate that $T_c$ is not found in the measured temperature range ($T>$1.4 K). The superconducting phase appears with the background of the Korringa law (1/$T_1T$ = const.). Antiferromagnetic fluctuation develops in the region displayed in bright  (yellow in color) region.  }
\label{f1}
\end{figure}

Figure 6 shows a new phase diagram for fcc Cs$_3$C$_{60}$. $T_N$ and $T_c$ vs unit cell volume at room temperature are plotted with the magnitude of 1/$T_1T$ indicated in brightness.  Superconductivity is only seen for the paramagnetic metal [dark (blue in color) region], where the Korringa law (1/$T_1T$ = const.) is well established in the entire temperature range. Antiferromagnetic fluctuations due to the localization of electron spins in C$_{60}$ balls are found in a limited region [bright (yellow in color) region]. The ground states of both antiferromagnetic insulator (Mott state) and superconductivity are completely separated. This phase diagram is fundamentally different from those of heavy-fermion compounds, high-$T_c$ copper oxides, two-dimensional organic conductors, and iron-pnictides. Namely, the superconductivity in fcc Cs$_3$C$_{60}$ breaks up at the Mott transition. Preventing the localization of electrons on C$_{60}$ balls is rather essential for Cooper pair formation and for finding still higher $T_c$ in this system.  

Although the Cs$_3$C$_{60}$ (T) phase was observed in previous works, we find for the first time that it does not become superconducting at $P$ = 4.7 kbar where the partial superconducting transition is observed. Furthermore, the observed phase separation into the Mott state and paramagnetic metal can reasonably explain the unresolved issues in Cs$_3$C$_{60}$, the reason (1) why shielding fraction depends on pressure and (2) why antiferromagnetism and superconductivity can "coexist" under pressure.

\section{Conclusions}
In this paper, we have presented a $^{133}$Cs-NMR study of face-centred cubic structured fulleride Cs$_3$C$_{60}$ under pressure. Different from previous reports, we find that Cs$_3$C$_{60}$ crystals show a phase separation between the Mott state and the paramagnetic metal. Our results indicate that the pressure-induced superconductivity in Cs$_3$C$_{60}$ is realized only for a paramagnetic metal. The phase diagram obtained from the present NMR study strongly indicates that the Mott state and superconductivity are competing orders.  The prevention of the localization of electron spins on C$_{60}$ balls is rather essential for finding further higher $T_c$ superconductivity in fullerides.

\begin{acknowledgments}

We thank Takashi Kambe for providing the Cs$_3$C$_{60}$ samples. S.K. thanks Y. Ihara and K. Ishida for useful discussion.
The X-ray diffraction patterns were partly measured in research projects (2007G612) of KEK-PF. This work was supported in part by research grants from MEXT (Nos. 22103004, 22013012, 22740232, and 23102717). 

\end{acknowledgments}

%\section{}\label{}
%\bibliography{apssamp}

\end{document}